\documentclass{revtex4}

\usepackage{graphicx}

\usepackage{amssymb}

\usepackage{psfrag}

\begin{document}

\title{Oscillatory fracture paths in thin elastic sheets}

\author{Beno\^{\i}t Roman, Pedro M. Reis, Basile Audoly,
Simon De Villiers, Vincent Vigui\'e, Denis Vallet }
\affiliation{PMMH, UMR 7636 CNRS/ESPCI, 10 rue Vauquelin 75231 cedex 5
Paris, France \\
 Manchester Center for Nonlinear Dynamics, Dept. physics and
astronomy, University
 of Manchester, M139PL UK \\
LMM, UMR 7607 CNRS/UPMC,  4 place Jussieu, case 162, 75252 Paris
cedex 05, France }

\begin{abstract}
We report a novel mode of oscillatory crack propagation when a
cutting tip is driven through a thin brittle polymer film.
Experiments show that the amplitude and wavelength of the
oscillatory crack paths scale linearly with the width of the
cutting tip over a wide range of length scales but are independent
of the width of the sheet and of the cutting speed.  A simple
geometric model is presented, which provides a simple but thorough
interpretation of the oscillations.
\texttt{http://www.lmm.jussieu.fr/platefracture/}

\end{abstract}

\maketitle

\label{sec:intro}
Despite a long history of research into the field of fracture, many
puzzling issues remain unsolved.  An interesting problem concerns the
direction of propagation of the crack tip: when a
glass breaks, can the shape of the resulting pieces be predicted?
Recent well controlled experiments have yield a variety of interesting
behaviours that are a challenge to existing theoretical formulations.
An oscillatory instability in dynamic cracks was recently observed in
a pre-tensioned thin rubber sheet~\cite{Deegan} whose mechanism is
still unclear.  Another example is the controllable quasi-static
propagation of oscillatory cracks in a thin strip of glass submitted
to a thermal field \cite{YS,Ronsin,Deegan2} which, despite its
apparent simplicity, has been stimulating much theoretical
study~\cite{Mokhtar,Ravi}.

\begin{figure}
    \psfrag{2}{$w=5\mathrm{~mm}$}
    \psfrag{3}{$w=31\mathrm{~mm}$}
    \psfrag{a}[Bc][Bl]{(a) experiments}
    \psfrag{b}[Bc][Bl]{(b) rescaled experiments}
    \psfrag{c}[Bc][Bl]{(c) simulation}
    \psfrag{x}[Bc][Bl]{$\times 6.2$}
    \psfrag{y}[Bc][Bl]{$\times 1$}
    \centering
    \includegraphics[width=.95\textwidth]{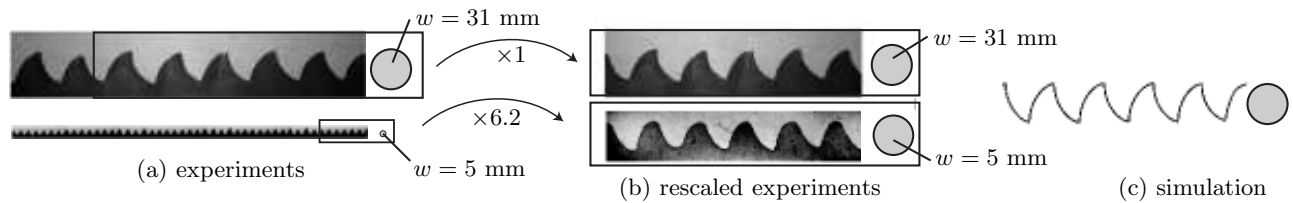}
    \caption{\emph{Right}: edge of the sheet seen from above
    (polypropylene $27\mathrm{~}\mu\mathrm{m}$ thick) for two
    diameters of the cylindrical cutting tip: $w=31$mm and $5$mm.  (a)
    Raw experimental snapshots of the cracked film edge and (b) after
    rescaling according to indenter radius; (c) numerical
    simulation.}\label{fig:oscl}
\end{figure}

Here we report novel results on oscillatory fracture paths in a new
experimental context: an object, which we denote by \emph{cutting
tip}, is perpendicularly driven through a thin polymer sheet held
along its lateral boundaries, and progressively cuts the material as
it advances.  For large enough cutting tip widths, the crack follows a
well defined and highly reproducible oscillatory path that spans a
wide range of length scales (Figure~\ref{fig:oscl}).  In fact, even doing
the experiment by hand yields surprisingly regular patterns.  The
experimental observations of oscillatory motion in this new geometry
present a challenge, from a fundamental point of view, to our
understanding of crack propagation in thin sheets.  Moreover, these
ideas should have practical applications since the precise cutting of
brittle thin sheets is common in industrial manufacturing.
\begin{figure}
    \psfrag{w}{$w$} \psfrag{D}{$D$}
    \psfrag{c}{clamp}
    \psfrag{s}{sheet}
    \psfrag{p}{$v$}
    \psfrag{S}{$S$}
    \psfrag{L}{$L$}
    \psfrag{I}{$I$}
    \psfrag{P}{$P$}
    \psfrag{u}{$\alpha$}
    \psfrag{v}{$\beta$}
    \psfrag{a}{$A=L\cup S$}
    \centering
    \includegraphics[width=5cm]{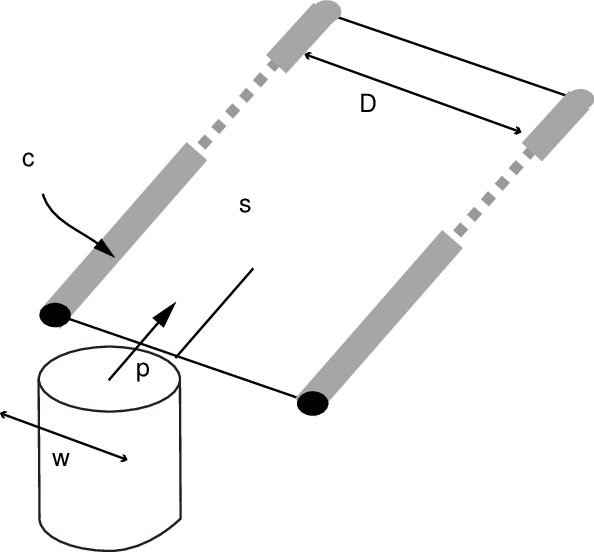}
    \includegraphics[width=.6\textwidth]{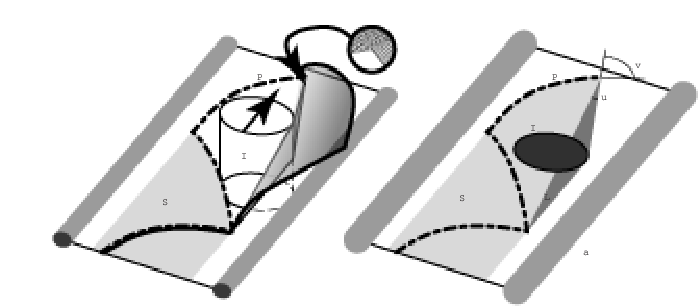}
    \caption{A cylindrical cutting tip is forced into a clamped sheet
    with a notch.  The cutting tip is displaced at constant speed $v$
    with respect to the plate, leading to an oscillatory crack path.
    Initial configuration of experimental setup (\emph{left}), typical
    configuration during fracture (\emph{center}) and geometrical 2D
    model (\emph{right}).  The strip is flexible in the smallest
    convex $S$ (light grey region) containing the crack path $P$, where
    it can easily bend without stretching.  Penetration of the cutting
    tip beyond this region induces tensiles stresses in the plate.
    Our simplified 2D geometrical model takes into acount this
    important feature through the angle $\alpha$ at the tip of the active
    zone $L$ shown in dark grey:  propagation is set to take place in mode
    $I\!I\!I$ loading (inset) along a direction given by $\beta$ when
    $\alpha$ exceeds a critical value $\alpha_{\mathrm{c}}$.}
    \label{fig:setup}
\end{figure}

We perform well controlled experiments with thin sheets of different
polymeric materials in a range of thicknesses and investigate the
dependence of the resulting fracture paths on the width of the cutting
tip.  We also propose a simple model that correctly reproduces the
behaviour of the fracture and illustrates the geometrical origin of
the oscillation.

\section*{Experimental oscillatory fracture paths}
A schematic diagram of the apparatus is presented in
Figure~\ref{fig:setup}.  It consists of a thin flat sheet (dimensions
ranging from $6\times 60\mathrm{~mm}$ to $12\times 50\mathrm{~cm}$)
clamped along its lateral boundaries and mounted on on a linear
translation stage.  This stage was driven at constant speed $v$
towards a fixed object, the \emph{cutting tip}, which could have
either rectangular or cylindrical profile with a variety of widths
($0.05\mathrm{~mm}<w<60\mathrm{~mm}$).  A camera was mounted directly
above the apparatus such that the propagating crack was imaged in the
cutting tip's frame of reference.

The sheet was initially prepared with a notch on one of its side
boundaries to position and initiate the crack.  Both bi-oriented
polypropylene and cellulose acetate thin sheets were investigated,
with thicknesses ranging between $25$ and
$130\mathrm{~}\mu\mathrm{m}$.  The sheet's Young's modulus and
fracture energy were measured to lie within the ranges
$E=1$--$2\mathrm{~GPa}$ and $\gamma=2$--$5\mathrm{~kJ}/\mathrm{m}^2$,
respectively.  Although they are made of polymers, these materials are
brittle as they were severely stretched when processed into thin
sheets.  As a result, they undergo minimal plastic deformation during
fracture propagation but, being thin, can sustain large bending without crack
initiation.  This explains why they are widely used in the packaging
industry (resistant but easy to tear once a notch is started).  The
oscillatory paths discussed below were not observed in ductile
materials.

As the thin sheet is forced through the fixed tip, the material is
cut, leaving behind a well defined and highly reproducible oscillatory
fracture pattern, some examples of which are shown in
Figure~\ref{fig:oscl}.  As one would expect, for thin enough objects
the path is straight.  The study of the transition from straight to
oscillatory paths will be presented elsewhere~\cite{art:lasuite} and
here we focus on the oscillatory regime for large cutting tips, that
is well above threshold.

\begin{figure}
    \psfrag{a}[c]{}
    \psfrag{b}[c]{}
    \psfrag{c}[c]{}
    \psfrag{d}[c]{}
    \psfrag{e}[c]{}
    \psfrag{f}[c]{}
    \psfrag{u}[c]{$\alpha$}
    \psfrag{v}[c]{$\beta$}
    \psfrag{x}[c]{$x$}
    \psfrag{y}[c]{$y$}
    \psfrag{A}[c]{(a)}
    \psfrag{B}[c]{(b)}
    \psfrag{C}[c]{(c)}
    \psfrag{D}[c]{(d)}
    \psfrag{E}[c]{(e)}
    \psfrag{F}[c]{(f)}
    \psfrag{N}{new contact}
    \centering
    \includegraphics[width=.95\textwidth]{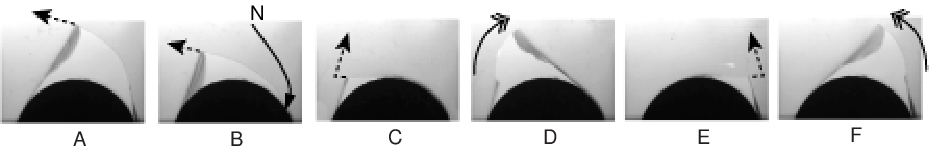}\\
    \includegraphics[width=.85\textwidth]{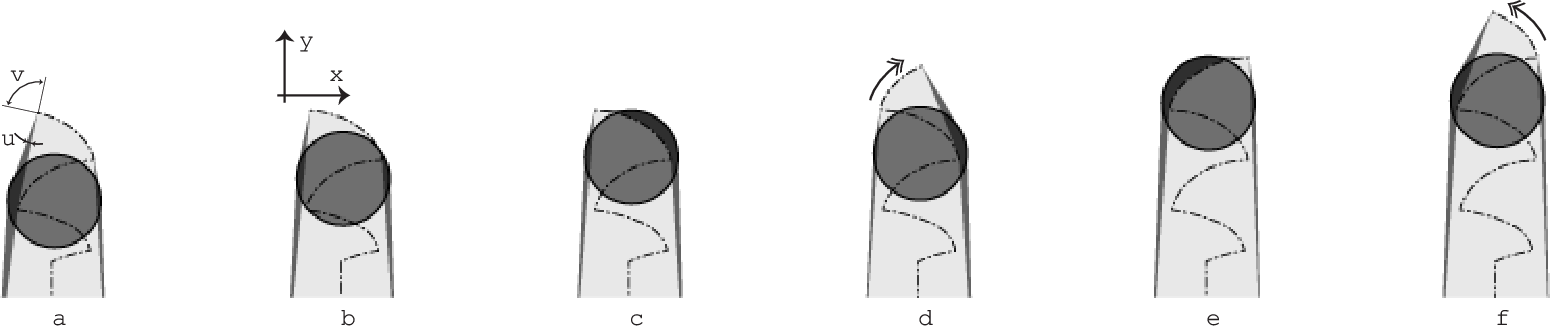}
    \caption{Similar sequential processes of the crack propagation
    over a single period (\emph{top:} experimental cutting tip in
    black; \emph{bottom:} corresponding frames in simulation with
    $\alpha = .153$ and $\beta = \pi/2$, crack path in dashed line).
    Frames (a--c): the left hand side of the crack edge is active and
    the crack tip advances leftwards.  In between (a) and (b), the
    cutting tip crosses the upper part of the crack curve and a new
    contact appears.  A kink then takes place, followed by a dynamical
    propagation (double arrow), both in the experiments and in the
    simulation, up to (d) which is the mirror image of the initial
    frame (a).  The second half-period up to (f) is similar.  In the
    simulation, the kink occurs because the active zone (dark grey
    area) is made up of two components, which rest on either side of
    the indenter.  Change of crack direction between (c) and (d)
    occurs when the angle $\alpha$ corresponding to the right hand
    component of the active zone becomes larger than that of the left
    hand side. Movies available at
    \texttt{http://www.lmm.jussieu.fr/platefracture}
    }
    \label{fig:expnum}
\end{figure}

Although these oscillatory patterns are reminiscent of those observed
in thermal quenching experiments~\cite{YS,Ronsin,Deegan2}, an
important difference is that the oscillation mechanism here arises
from a coupling with the large out-of-plane deflection of the thin
sheet, as seen in the experimental frames in Figure~\ref{fig:expnum}.

\section*{A simple geometrical model}
We believe that the origin of the oscillations in our experiment lies
in the mechanical properties of thin elastic sheets, and in their
connection with the geometry of surfaces.  Under external constraints,
thin elastic sheets bend in order to avoid stretching, since bending
energy is small compared to stretching energy~\cite{pogorelov}.  In
order to conserve their in-plane lengths and remain unstretched, thin
sheets tend to adopt developable shapes, that are the reunion of
straight lines, \emph{generatrices}.  Because the lateral edges of the
plate are held, these generatrices may not cross the clamped
boundaries of the plate and can only end up on the crack path.  We
thus define a ``\emph{soft}'' zone in the plane of the elastic plate
by considering the reunion of all the segments ending on the crack
path.  This is called the convex hull $S$ of the crack path (see
Fig.~\ref{fig:setup}, right).  In $S$, the sheet can deform by pure bending
and can therefore easily accomodate the presence of the cutting tip.
However, when the cutting tip penetrates beyond $S$, in a region which
we call ``\emph{hard}'', the sheet undergoes stretching which
eventually leads to crack propagation.


In order to determine the direction of propagation in our thin sheets
we have performed a complementary ripping (mode $I\!I\!I$) experiment
by systematically curving and pulling one side of an initial notch.
Crack propagation was observed to always follow a well defined and
highly reproducible angle $\beta$ with respect to the generatrix of
the fold.  This angle $\beta$ depends on the material and, since these
materials are anisotropic, on the orientation of the ripping, but we
consistently measured an angle close to $\pi/2$ to within 20\%.  So, to
first approximation, we take $\beta=\pi/2$.

The propagation criterion in our model is the following.  We consider
the ``large'' hull $A$, defined as the convex hull of the crack tip
\emph{plus} the horizontal section of cutting object (black disk in
the Fig.~2c) altogether.  The so called active area $L$ (in dark grey)
computed as the set difference $L=A\setminus S$ of the large and the
soft hull is the region of the hard zone where penetration
of the cutting tip induces tensile stresses.  When this ``active''
area includes a cone with angle $\alpha>\alpha_\mathrm{c}$ at the
crack tip greater than a critical value (see Fig.~\ref{fig:setup}c),
the crack is set to propagate instantaneously in the direction given
by the angle $\beta$, until $\alpha$ decreases back to
$\alpha_\mathrm{c}$ (note that $\alpha$ indeed decreases during crack
advance since the soft zone expands).  The angle $\alpha_\mathrm{c}$
results from a balance between in-plane stresses in the plate and the
energy needed to advance the crack: it can in fact be computed from
the mechanical and fracture properties of the film.  We refer to
Ref.~\cite{art:lasuite} for a more detailed presentation of this
geometrical fracture model and its rational derivation from elementary
principles of fracture and plate mechanics.

\section*{Experimental and numerical results}
A numerical simulation of this geometrical model yields oscillatory
crack paths over a wide range of values of the parameters $\alpha_c$
and $\beta$.  In Fig.~\ref{fig:oscl} we compare typical computed
(right) and experimental (left) crack paths.  The non-sinusoidal crack
patterns are in remarkably good agreement.  Their wavelength and
amplitude depend weakly on the parameters $\alpha_\mathrm{c}$ and
$\beta$ of the model and their overall shape is almost independent of
them.

The sequential processes of the crack propagation over a single period
is presented and discussed in Figure~\ref{fig:expnum} for both the
experiments and numerics.  The numerical simulations were run with
parameters $\alpha = .153$ and $\beta = \pi/2$.  The evolution of the
crack paths demonstrates strong similarities between the experiments
and the simulation based on our model.

\begin{figure}
    \psfrag{a}[c]{\small amplitude ($\mathrm{mm}$)}
    \psfrag{l}[c]{\small wavelength ($\mathrm{mm}$)}
    \psfrag{w}[c]{\small width $w$ (mm)}
    \centering
    \includegraphics[scale=.4]{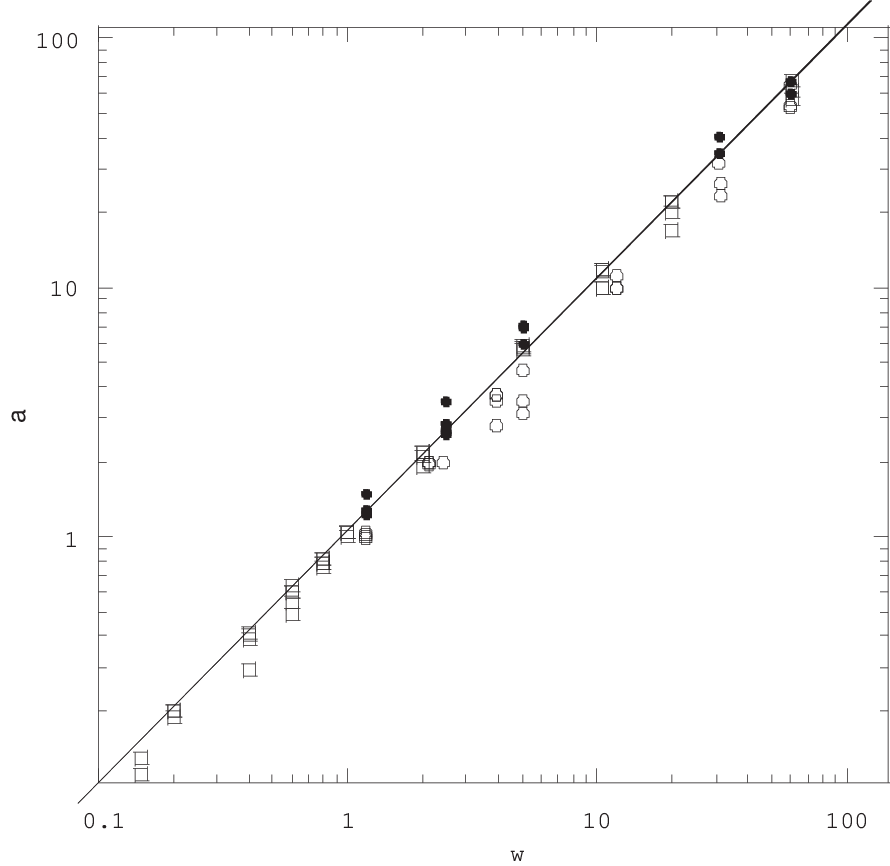}
    \hspace{.2\textwidth}
    \includegraphics[scale=.4]{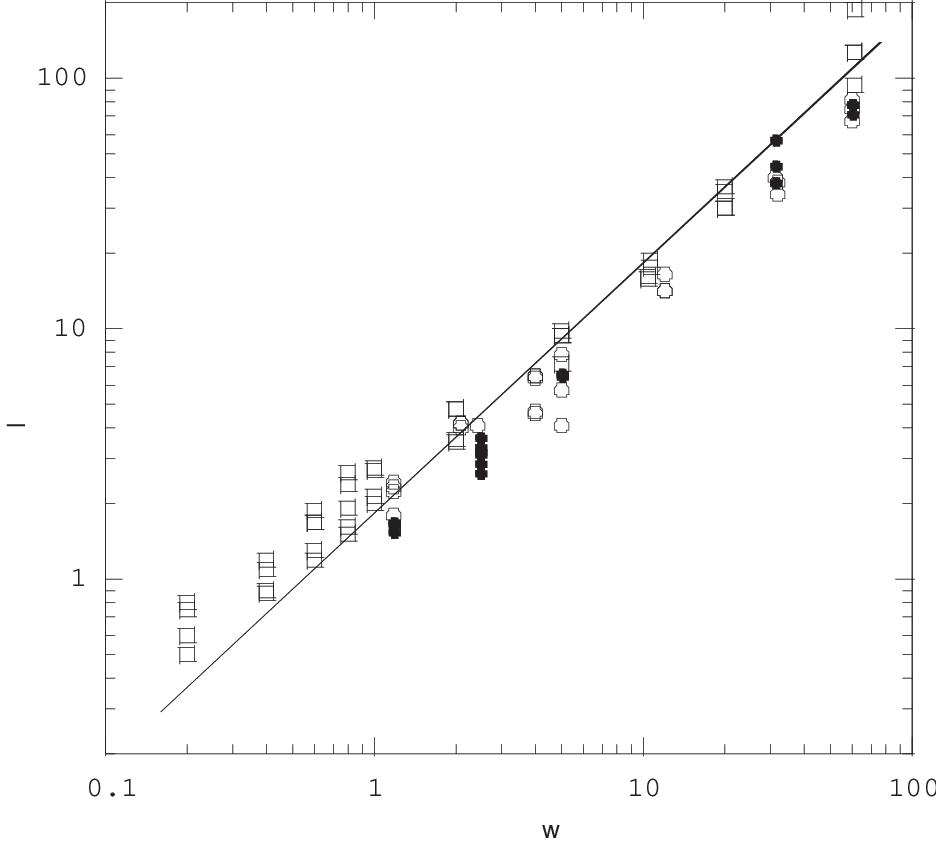}
    \caption{Log-log plots of the peak-peak amplitude and wavelength
    of the oscillations as a function of tip width $w$ for various
    materials and thicknesses (polypropylene 25 to 53 $\mu$m: open
    symbols, cellulose acetate 100 and 130 $\mu$m : closed symbols),
    and for both cylinders ($\circ \bullet$) and rectangular blades
    ($\Box$) : amplitude and wavelength organize on slope 1 lines, on
    a range of almost 3 decades.  The straight lines represent the
    best linear fits with slope 1.}
    \label{fig:amplitude}
\end{figure}

Our geometrical model includes a single length scale, the size of the
cutting tip, and does not take into account its driving speed.  In fact, we
have performed experiments for
$0.06\mathrm{~mm}/\mathrm{s}<v<64\mathrm{~mm}/\mathrm{s}$ and found no
influence of speed of the patterns.  We have also checked
experimentally, for a fixed tip size $w=2\mathrm{~mm}$, the pattern to
be independent of the sheet's width, $D$, within the range $3.3w < D <
21w$, in agreement with the predictions of the model.  This feature is
\emph{a priori} surprising since one could have thought of $D$ as one
of the natural length scales.  Having determined that there is only
one length scale $w$ in the problem, one expects both the wavelength and
amplitude of the oscillations to scale linearly with it.
Indeed, as shown on Fig.~\ref{fig:amplitude}, we experimentally find a
linear dependence of the amplitude and wavelength on the tip width
over almost three decades, for a variety of polymeric materials and
thicknesses. This supports a purely geometric fracture mechanism in
this regime.
Note that this linear behaviour breaks down in the vicinity of the transition
region between straight and oscillatory paths: there, the sheet can no longer
be considered thin relative to the cutting tip and the model's
assumptions are not valid anymore (the plate bending and stretching
energies become of the same order of magnitude).

To conclude, we have reported a novel mode of oscillatory crack
propagation as a cutting tip of moderately large width is driven
through a thin polymer sheet.  We proposed a geometric model for these
cracks, which retains the salient mechanical properties of thin
elastic sheets: our scenario is based on the coupling between the
direction of crack propagation and large deflections of the plate.
Although very simple, such a model reproduces surprisingly many
experimental features of the cracks, including their shape and
dynamics.  The crack path amplitude and wavelength were found
independent of the driving speed, of the thickness and lateral width
of the sheet, and scale linearly with the size of the object, in
agreement with the model.  This explains why this phenomenon is highly
robust and insensitive to irregularities in the speed and boundary
conditions, and can even be observed if the cutting is done by hand.




The authors wish to thank Mokhtar Adda Bedia for stimulating discussion,
and Charles Baroud for putting in contact the Paris and Manchester
groups who started working on this subject independently.



\begin{thebibliography}{00}


\bibitem{YS} Yuse A., Sano M. , Nature (London) 362 (1993) 329--331 .

\bibitem{Ronsin} Ronsin O., Heslot F.and Perrin B., experimental study of
quasistatic brittle crack propagation Phys. Rev. Let. 75 (1995)

\bibitem{Mokhtar} Adda-Bedia  M. and Pomeau  Y., Crack instabilities of a heated
glass strip, Phys. Rev. E 52 (1995) 4105--4113.

\bibitem{Ravi} Yang B. \& Ravi-Chandar K., crack path instability in a quenched
glass plate, J. Mech. Phys. Solids 49 (2001) 91--130

\bibitem{Deegan2} Deegan R., Chheda S., Patel L., Marder M., Swinney H.,
 Wavy and rough crack in Silicon, 2002 preprint

\bibitem{Deegan} Deegan R., Petersan P., Marder M., Swinney H.,
Oscillating fracture paths in rubber Phys. Rev. Let. 88 (2002)


\bibitem{pogorelov} Pogorelov A., Bending of surfaces and stability of
shells, Translation of mathematical monographs. AMS, 1988.

\bibitem{art:lasuite} B. Audoly, P. M. Reis, B. Roman, D. Vallet, V.
ViguiŽ, S. DeVilliers, in preparation.





\end{thebibliography}
\end{document}